# The Brazilian Tunable Filter Imager for the SOAR telescope

Cláudia Mendes de Oliveira[1], Keith Taylor[1], Bruno Quint[1], Denis Andrade[2], Fabrício Ferrari[3], Rene Laporte[4], Giseli de A. Ramos[5], Christian Dani Guzman[6], Luiz Cavalcanti[2], Alvaro de Calasans[7], Javier Ramirez Fernandez[2], Edna Carolina Gutierrez Castañeda[1], Damien Jones[8], Fernando Luis Fontes[9], Ana Maria Molina[1], Fábio Fialho[2], Henri Plana[10], Francisco J. Jablonski[4], Luiz Reitano[4], Olivier Daigle[11], Sergio Scarano Jr.[12], Philippe Amram[13], Philippe Balard[13,] Jean-Luc Gach[13], Claude Carignan[14]

[1]Departamento de Astronomia, Instituto de Astronomia, Geofísica e Ciências Atmosféricas da Universidade de São Paulo, Rua do Matão 1226, Cidade Universitária, 05508-090, São Paulo, Brazil

[2]Escola Politécnica da Universidade de São Paulo, Av. Prof. Luciano Gualberto, travessa 3, no. 380, Cidade Universitária, 05508-010, São Paulo, Brazil

[3]Instituto de Matemática, Estatística e Física, FURG, Av. Itália, Km 8, 96201-900, Rio Grande, RS, Brazil

[4]Instituto Nacional de Pesquisas Espaciais/MCT, Av. dos Astronautas, 1758, 12227-010, São José dos Campos, SP, Brazil

[5]Instituto de Matemática e Estatística da Universidade de São Paulo, Rua do Matão 1210, Cidade Universitária, 05508-090, São Paulo, Brazil

[6]Departamento de Engenharia Eletrica, Pontificia Universidad Católica de Chile, Vicuña Mackenna 4860, Casilla 306, 22 Santiago, Chile

[7]Instituto de Física, Universidade de São Paulo, CP 66318, 05314-970, São Paulo, Brazil

[8]Prime Optics, 17 Crescent Road, Eumundi, Qld 4562, Australia

[9]Universidade Paulista - Unip, Rod. Presidente Dutra, km 157,5, Pista Sul, CEP 12240-420 São José dos Campos, São Paulo, Brazil.

[10]Laboratorio de Astrofísica Teórica e Observacional, Universidade Estadual de Santa Cruz, Rodovia Ilhéus-Itabuna km 16 45650-000 Ilhéus BA, Brazil

[11]Nüvü Caméras, Montréal, Qc, Canada

[12]Southern Astrophysical Research Telescope (SOAR), Casilla 603, La Serena, Chile

[13]Laboratoire d'Astrophysique de Marseille, Université de Provence, CNRS, 38, rue Frédéric Joliot-Curie 13388, Marseille cedex 13, France

[14]Laboratoire d'Astrophysique Expérimentale, Observatoire du Mont Mégantic, and Département de Physique, Université de Montréal, CP 6128, Succursale Centre-Ville, Montréal, PQ H3C 3J7, Canada

# ABSTRACT

This paper presents a description of a new Tunable Filter Instrument for the SOAR telescope. The Brazilian Tunable Filter Imager (BTFI) is a highly versatile, new technology, tunable optical imager to be used both in seeing-limited mode and at higher spatial fidelity using the SAM Ground-Layer Adaptive Optics facility (SOAR Adaptive Module) which is being deployed at the SOAR telescope. Such an instrument opens important new science capabilities for the SOAR astronomical community, from studies of the centers of nearby galaxies and the insterstellar medium to statistical cosmological investigations.

The BTFI concept takes advantage of three new technologies. The imaging Bragg Tunable Filter (iBTF) concept utilizes Volume Phase Holographic Gratings in a double-pass configuration, as a tunable filter, while a new Fabry-Perot (FP) concept involves the use of commercially available technologies which allow a single FP etalon to act over a very large range of interference orders and hence spectral resolutions. Both these filter technologies will be used in the same instrument. The combination allows for highly versatile capabilities. Spectral resolutions spanning the range between 25 and 30,000 can be achieved in the same instrument through the use of iBTF at low resolution and scanning FPs beyond R ~2,000 with some overlap in the mid-range. The third component of the new technologies deployed in BTFI is the use of EMCCDs which allow for rapid and cyclically wavelength scanning thus mitigating the damaging effect of atmospheric variability through the acquisition of the data cube.

An additional important feature of the instrument is that it has two optical channels which allow for the simultaneous recording of the narrow-band, filtered image with the remaining (complementary) broad-band light. This then avoids the otherwise inevitable uncertainties inherent in tunable filter imaging using a single detector which is subject to temporal variability of the atmospheric conditions.

The system was designed to supply tunable filter imaging with a field-of-view of 3 arcminutes on a side, sampled at 0.12" for direct Nasmyth seeing-limited area spectroscopy and for SAM's visitor instrument port for GLAO-fed area spectroscopy. The instrument has seen first light, mounted on the SOAR telescope, as a visitor instrument. It is now in comissioning phase.

1. **Introduction**

The BTFI concept arose as a response to the need within the Brazilian community for a 3D-spectroscopy instrument able to make use of SAM's full field of view (SAM is the SOAR's ground-layer adaptive optics instrument, Tokovinin et al. 2008). SAM's potential for giving enhanced spatial resolution over a relatively large (~3 x 3 arcmin²) field of view was always envisaged to be a powerful tool that would allow the SOAR community to conduct high impact scientific programs. In order to fully realize the science potential that such an investment allows, it was necessary to utilize not just the superb image quality but also the field-of-view advantage of SAM for not only imaging but also spectroscopy. SOAR already possesses an optical imager (SOI, the SOAR Optical Imager) at the bent Cassegrain focus. Furthermore, SAM has planned the construction of a dedicated GLAO-enhanced optical imager (SAMI, the SOAR Adaptive Module Imager). On the other hand, the BTFI project represents the development of a wide-field tunable filter imager as an effective means for performing area spectroscopy over a wide range of spectral resolving powers on both the SOAR Nasmyth focus and on SAM, so as to fully exploit their science potential.

Of special interest to the Brazilian community is the study of the centers of nearby active galaxies, the study of kinematics and metallicities of cluster and group galaxies at redshifts 0.1-0.3 (for which a number of systems can be observed in one shot) and of stellar mass loss phenomena in the surrounding interstellar medium. For these studies, a larger field of view than the one delivered by SIFS (the SOAR Integral Field Spectrograph, with a field of view of 3 x 5 arcsec, also to be used with SAM, Lepine et al. 2003), was desirable. Moreover, there is currently no Fabry-Perot, Tunable Filter instrument on any telescope working with adaptive optics. It is therefore clearly recognized within the Brazilian community that BTFI offers new capabilities that are worth exploring.

The BTFI project started in February 2007 and it successfully passed its CoDR and PDR in September 2008 and June 2009 respectively. The system was designed to supply tunable filter imaging with a field-of-view of ~3' sampled at ~0.12" (with an f/7 camera) for direct Nasmyth seeing-limited area spectroscopy and for SAM's Visitor Instrument port for GLAO-fed area spectroscopy.

Like many other instruments of its type, BTFI employs Fabry-Perots (FPs) in order to achieve high spectral resolutions up to R ~30,000. In the less explored, low spectral resolution domain, exploited more recently by the Anglo-Australian Telescope's TAURUS Tunable Filter (TTF), the BTFI will utilize a new double-pass Volume Phase Holographic (VPH) grating technology (the imaging Bragg Tunable Filter) to achieve ultra-low to intermediate ($25 < R < 4000$) spectral resolving powers in a highly efficient, cost-effective and compact configuration.

The instrument is being developed by the Instituto de Astronomia, Geofísica e Ciências Atmosféricas (IAG) at the Universidade de São Paulo, Brazil, in collaboration with several other Brazilian Institutions, such as Escola Politécnica (POLI) from the same university, Instituto Nacional de Pesquisas Espaciais (INPE), Laboratório Nacional de Astrofísica (LNA) and Universidade Federal do Rio Grande, Universidade Estadual De Santa Cruz and international collaborations with the Laboratoire d'Astrophysique de Marseille (LAM), the University of Montreal and Universidad Catolica in Chile.

The paper is organized as follows: in Section 2 a description of previous instruments of similar type is presented while in section 3 we outline the instrument concept and its new technologies. In Section 4 we briefly describe the science cases. In Section 5 we detail the BTFI instrument itself and we present the first observations obtained in January 2012.

2. **Fabry-Perot and Tunable Filter interferometers**

In order to clarify the concept of spectroscopy, following Fellgett (1958), two classes should be defined: the spectrographs and the spectrometers. A spectrograph allows a spatial measurement of the position of maximum intensity of a line or of a fringe on the detector while a spectrometer allows its temporal measurement. In other words, a spectrograph is associated to an unique image obtained during an unique reading of the image sensor while a spectrometer is associated to a scanning sequence obtained during several readings of the detector. Both spectroscopies are equally efficient if, at any time, the whole surface of the sensor is optimally used. Spectrographs usually use grisms to disperse the light while spectrometers require interferometers. Examples of spectrometers are Fabry-Perots (FPs) or Tunable Filters (TFs; or low spectral resolution Fabry Perots are sometimes called tunable filters) and Fourier transform imager systems (FTSs). Their "*ecologic*" niches are different; spectrographs are used when the science drivers request large spectral ranges and small FoVs while spectrometers are preferred when larger FoVs and smaller spectral ranges are preferred.

Providing a 2D-image within a given spectral band using a Fabry-Perot instrument requires scanning the interferometer. Following the Fabry-Perot interference formula, *2ne cos(i) = pλ* (where *n* in the refractive index, *e* the distance between the two parallel plates, *i* the angle of incidence of the light, *λ* the wavelength and *p* the interference order), the scan can be achieved by changing *i* (selection of the angle on the sky), *n* (the index of the layer between the plates, usually through varying the gas pressure between the plates of the interferometer) or *e*, by moving the distance between the two plates. Modern interferometers have generally chosen to scan by acting on *e* but this was not the case for the first Fabry Perot imagers. It is beyond the scope of this section to provide a historical review of the Imaging Fabry-Perot systems. However we should nevertheless mention the pioneer work of Courtes (1960) in Marseille who scanned the field of view by varying the angle *i*, a technique used later by other groups (e.g. de Vaucouleurs & Pence, 1980). Scanning Fabry-Perots through changing *n* (pressure) were developed at Maryland by Tully (1974); Roesler et al (1982) for the instrument PEPSIOS, Smith (1981) for the instrument SPIFI and at Rutgers by Williams, Caldwell & Schommer (1984). Scanning by changing the gap was pioneered by Taylor & Atherton (1980) in the instrument

TAURUS followed by Boulesteix et al (1984) with the instrument CIGALE and SPIFI and Rutgers groups.

In the following section we will quickly tabulate some of the more recent Fabry-Perot spectrometers available for astronomy, all of them scanning the gap e. All of the Fabry-Perot interferometers on large telescopes fit within the collimated section of a focal reducer, following the pioneer concepts suggested by Courtès (1960). All of them are also seeing limited with a spatial sampling that depends on the detector pixel scale. Their spectral resolutions only depend on the interference order, p, of the interferometers (for a given reflective factor R of the plates at a given wavelength), furthermore the spectral resolutions given in Table 1 are those usually used with the instruments, generally with different Fabry-Perot etalons.

**TAURUS & TTF** - Taurus is an imaging Fabry-Perot interferometer which was used at the AAT between 1981 and 1983 (Taylor & Atherton, 1980; Atherton et al. 1982). Taurus was the first scanning imaging Fabry-Perot in use for astronomy. The Taurus-II Tunable Filter (TTF) was a more powerful version of the original Taurus; it was in regular use from 1996 to 2003 on the Anglo Australian Telescope (AAT). During this period, a duplicate was also used on the William Herschel Telescope (WHT) from 1996 to 2000. An important feature of the TTF was the use of charge shuffling synchronized to band switching in order to greatly suppress systematic errors associated with conventional imaging (Bland-Hawthorn & and Jones, 1998a; Bland-Hawthorn & Jones, 1998b). Taurus-II is no longer offered or supported at the AAT or WHT.

**Table 1: Instrumental Parameters of the Fabry-Perot interferometers.**

| Name[1] | Status[2] | Telescope[3] | Wv Range[nm][4] | $R_\lambda$[5] | FoV[ ' ][6] | S[ " ][7] |
|---|---|---|---|---|---|---|
| TAURUS | Out. Op. | AAT(3.89m) & WHT (4.2m) | 370 – 950 | 100 - 60000 | 9.87 | 0.37 |
| HIFI | Out Op. | CFHT(3.58m) & UH(2.2m) | 400-750 | 4000-16000 | 10 | 0.43-0.69 |
| CIGALE | Out Op. | ESO (3.60m) | 656.3 – 678.2 | 15000 | 5.0 | 0.45 |
| PALILA | Out Op. | CFHT(3.58m) | 656.3 – 678.2 | 15000 | 5.8 | 0.34 |
| MOS-FP | Out Op. | CFHT(3.58m) | 365 – 1000 | 5000 – 15000 | 10 | 0.8 |
| GriF | Out Op. | CFHT(3.58m) | H & K Band | 2000 | 0.6 | 0.12 |
| GHASP | In Op. | OHP(1.92m) | 656.3 – 678.2 | 15000 | 5.8 | 0.68 |
| GHaFaS | In Op. | WHT(4.2m) | 656.3 – 678.2 | 5000 – 15000 | 4.0 | 0.45 |
| FaNTOmM | In Op. | Mégantic(1.6m) | 656 – 678 | 5000 – 15000 | 19.4 | 1.61 |
| PUMA | In Op. | San Pedro(2.1m) | 365 – 865 | 10650 | 10 | 0.67 |
| SCORPIO | In Op. | SAO(6m) | 500 – 900 | 3000-10000 | 6.1 | 0.40 |
| MMTF | In Op. | Magellan(6.5m) | 500 – 920 | 200-1840 | 27 / 10 | 0.60 |
| RSS/FP | In Op. | SALT(11m) | 430 – 860 | 300 – 9000 | 8 | 0.25 |
| OSIRIS | Future | GRANTECAN (10.4m) | 365 – 1050 | 300 – 5000 | 7.8 | 0.13 |
| TFI | Future | JWST(6.5m) | 1500 – 5000 | 75 – 120 | 2.2 | 0.60 |

[1] Name. [2] Present status. Out Op.: Out of Operation and In Op. In Op.: In Operation. [3] Telescope(s) and primary mirror size. [4] Wavelength range. [5] Spectral resolution. [6] Field of view. [7] Spatial sampling (pixel size).

**HIFI** – (Hawaii Imaging Fabry-Perot Interferometer) was a low resolution Fabry-Perot Imager that differed from the TAURUS systems in its use of large free spectral-range etalons (~ 100Å) with high finesse (~ 60) (Bland & Tully 1989) with a CCD at its image plane. HIFI was used both on the University of Hawaii 2.2m and CFHT 3.6 telescopes. It provided seeing limited observations.

**CIGALE** - CIGALE (for CInematics of GALaxiEs) is an imaging Fabry-Perot interferometer built by the Observatoire de Marseille (Boulesteix et al. 1984). It was used on several telescopes: CFHT, the 2.6m Byurakian Telescope, the 6m Zelenchuk Telescope and the 3.6m ESO telescope. It is composed of a focal reducer, a scanning Fabry-Perot and an Image Photon Counting System (IPCS). The IPCS, with a time sampling of 1/50 second and zero readout noise, makes it possible to scan the interferometer rapidly (typically 5 seconds per channel), avoiding sky transparency, air-mass and seeing variation problems during the exposures.

**PALILA** - The focal reducer PALILA was built by the Observatoire de Marseille for the CFHT and was in use at CFHT from 1990 to 1994 until it was donated to the Observatoire du mont Mégantic. It differs from CIGALE in that the detector was a CCD camera instead of an IPCS (Boulesteix & Grundseth, 1987)

**MOS-FP** - MOS/SIS was a dual Multi-Object and Subarcsecond Imaging Spectrograph for CFHT which contained a Fabry-Perot facility (MOS-FP). MOS-FP replaced PALILA in 1994. The MOS/SIS spectrograph was jointly designed and built by teams from the Dominion Astrophysical Observatory in Victoria, the Observatoire de Paris-Meudon, the Observatoire de Marseille and CFHT. MOS-FP saw its first light in July 1992 and it has not been used any longer since 2006. (Le Fèvre et al, 1994)

**GriF** - The three-dimensional spectroscope GriF offered Fabry-Perot capabilities in the near-infrared behind PUEO, the CFHT adaptive optics bonnette, and provided images at the diffraction limit of the telescope in the K band (Clénet et al., 2001). GriF is no longer offered at CFHT.

**GHαSP** - (for Gassendi Hα survey of SPirals) is a CIGALE-like instrument attached to the Cassegrain focus of the 1.93m telescope at the Observatoire de Haute-Provence equipped with a scanning Fabry-Perot and a photon counting detector. The instrument has been in continuous operation since 1998. (Garrido et al 2002)

**GHαFaS** - (for Galaxy Hα Fabry-Perot System for WHT) is a Fabry-Perot system available at the William Herschel Telescope. It was mounted, for the first time, at the Nasmyth focus of the 4.2m WHT in La Palma in July 2007. With a spectral resolution of the order R ~15,000 and a seeing limited spatial resolution, GHαFaS provides a new view of the Ha-emitting gas over a 4 arcminutes circular field in the nearby universe (Carignan et al. 2008)

**FaNTOmM** - (for Fabry-Perot de Nouvelle Technologie pour l'Observatoire du mont Mégantic) is the combination of a focal reducer (PANORAMIX: the 1.6m mont Mégantic OmM telescope focal reducer), a scanning Fabry Perot and an IPCS. FaNTOmM is a third generation instrument using a photon counting camera (IPCS) based on an GaAs photo cathode that can achieve quantum efficiency of up to 28%, comparable to a thick CCD, but with zero readout noise (Hernandez et al. 2003).

**PUMA** - (The UNAM Scanning Fabry-Perot Interferometer) is an integral field spectrometer having as the dispersive element a scanning Fabry-Perot interferometer working at optical wavelengths optimized in the red (Rosado et al. 1995). The instrument is attached to the 2.1m telescope of the Observatorio Astronomico Nacional at San Pedro Martir, B.C., Mexico.

**SCORPIO** - (Spectral Camera with Optical Reducer for Photometrical and Interferometrical Observations) is a multi-mode focal reducer containing a Fabry-Perot facility (Afanasiev & Moiseev 2005). 2D spectroscopic observations using an imaging Fabry-Perot at the 6m telescope was initiated in the early 80s using the CIGALE system. In 1997 a CCD was attached to the old focal reducer instead of a photon counter. A new multi-mode focal reducer SCORPIO was developed by the Special Astrophysical Observatory, Russian Academy of Sciences, and was first on sky in 2000.

**MMTF** - The Maryland Magellan Tunable Filter on the Magellan-Baade Telescope is a narrowband filter which is tunable in both central wavelength and transmission band-pass. It has a large field of view (27' full diameter of which ~10' is monochromatic) which means that if the target covers a range in velocity larger than the 10' of the central spot (depending on the wavelength), multiple exposures are needed to capture all line emission. The MMTF operates on similar principles to the Taurus Tunable Filter (Veilleux, S., et al. 2010).

**RSS** - (Robert Stobie Spectrograph) is designed and built for SALT (Southern African Large Telescope) and has the capability to obtain true wide-field imaging spectroscopy through its Fabry-Perot (FP) modes (Rangwala et al. 2008). It has been progressively put in operation since 2010.

**OSIRIS** - (Optical System for Imaging and low-intermediate Resolution) is an imager and spectrograph for the optical wavelength range, located at the Nasmyth-B focus of GTC (Cepa et al. 1998). It provides narrow-band tunable filter imaging and charge-shuffling capabilities. The blue Tunable Filter mode of OSIRIS has been delivered and is undergoing technical verification tests on the telescope. The on-sky commissioning will be done during 2012.

**TFI** - (Tunable Filter Imager) in an uncertain science instrument for JWST. It is a sensitive camera that shares the optical bench of the Fine Guidance Sensor. The TFI can also perform imaging with a choice of four coronagraphs as well as a non-redundant mask (Ingraham et al. 2010).

3. **Instrument concept – the new technologies used in BTFI**

The concept of a classical FP-based imaging interferometer for both kinematic work (high interference order) and tunable filter work (low interference order) is depicted in Figure 1. This will later be contrasted with the iBTF technology, however for now it illustrates the basic concept of 3D data cubes for both techniques. The BTFI instrument incorporates both technologies (FP + iBTF) in order to give it great versatility for a wide range of new science.

### 3.1. The Fabry-Perot operating modes of BTFI

BTFI uses two FP etalons that can be used individually or in tandem. The FP is typically mounted in the collimated beam for high order of interference, high spectral resolution work or in the divergent beam, near the input focus, for low order of interference low resolution work or in tandem where the low order FP can be used as a tunable order blocker for the high resolution etalon. Each of these modes will be described in detail below.

### 3.1.1. The classical Fabry-Perot mode

The raw data produced by a scanning FP can be represented by a series of images of the studied object, obtained at different wavelengths (or radial velocities) emanating from the source. The different wavelengths are obtained by changing the spacing between the plates of the FP etalon. The observed wavelength range is isolated using an interference filter. Once calibrated, one has a spectrum for each pixel in the field. Such FP systems have been used primarily for two types of application:

  i) To obtain precise line profiles ($R > 25{,}000$) in order to derive the physical parameters of emission-line regions.
  ii) To obtain the complete 2-D kinematics of an emission-line source.

While in the nearby universe, the H$\alpha$ line is mainly used, it can be replaced by the [O II], H$\beta$ or [O III] lines for higher redshift galaxies. Broadly, the science requires the scanning FP mode of the BTFI to possess the following characteristics:

- Wavelength range: 400 – 1,000nm
- Resolution: up to 30,000
- Pixel size of 15 microns (~0.12" on the sky)
- Field of view: 3 arcmin on a side

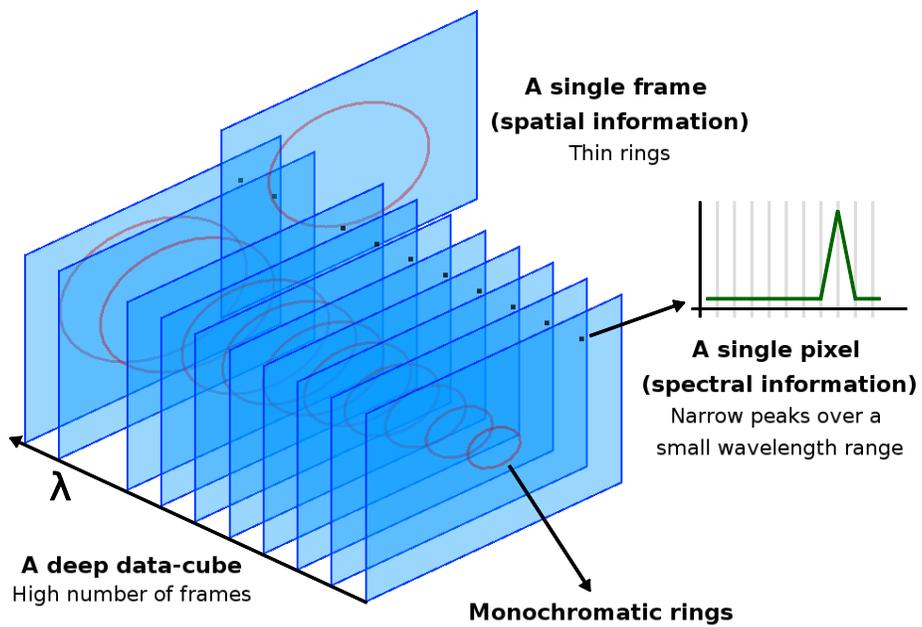

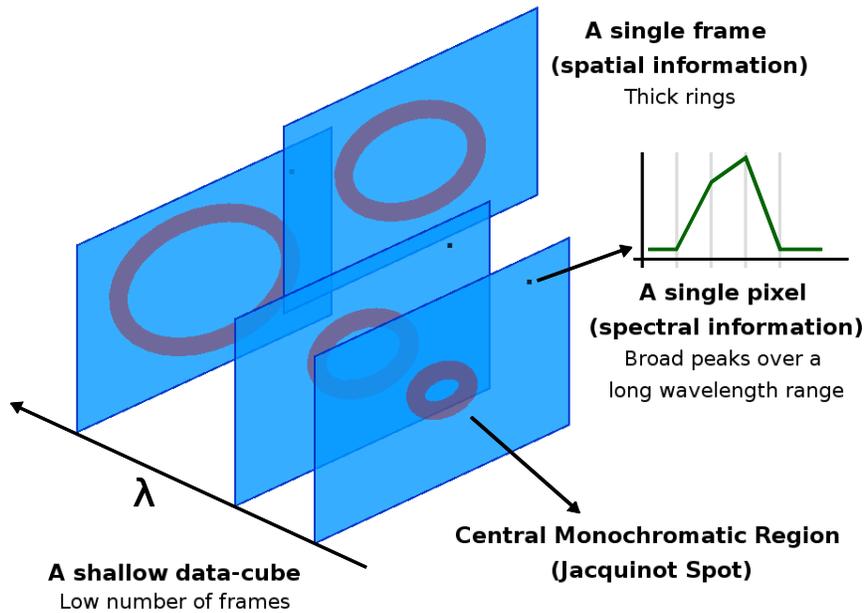

**Figure 1: Illustration of how a Fabry-Perot 3D Spectrograph works. The top figure shows a high interference order FP that yields a stack of high resolution images within a narrow spectral range. The bottom figure represents a low interference order FP and the resulting stack of low resolution images with a larger spectral range.**

For BTFI, in the classical FP mode, we use an FP etalon mounted in the collimated space. This FP etalon was manufactured by SESO (Société Européenne de Systèmes Optiques - http://www.seso.com/) to allow a far greater range of spectral resolutions than available using the more traditional designs. Older style Fabry-Perots can only cover a few free spectral ranges due to the limited scan range of their piezo-electric transducers. The technology developed for the new SESO etalon by the Cedrat company (http://www.cedrat.com), allows us to enlarge the scan range of the piezo-electric transducer from just above zero up to ~250μm, allowing us to cover hundreds of orders. With BTFI we would target a resolution range for the classical FP mode between R~6,000 and R~30,000.

### 3.1.2. The FP Tunable Filter mode

The first FP Tunable Filter in regular use on a 4m telescope was the Taurus Tunable Filter (TTF), commissioned fully on the AAT in 1995-97 (Bland-Hawthorn and Jones 1998b) The TTF employed a pair of low resolution FP etalons covering 370-650nm (blue range) and 650-960nm (red range) which were deployed separately. The FP etalons for the TTF were manufactured by Queensgate Instruments Ltd. and were conventional etalons other than having a small nominal gap (low order interference). Unlike conventional Queensgate etalons which had been used in the previous generation of FP interferometers, the TTF also incorporated long-range piezo-electric stacks (which expand the otherwise available plate separation range) and high performance coatings covering almost half the optical wavelength range. Since FPs have a periodic transmission profile, the instrument requires a limited number of order blocking filters. At low resolution (R ~300), conventional broad-band UBVRIz filters suffice. At the higher resolution end of the range (R = 1,000), eight intermediate-band filters were used. Thus the TTF system was used to obtain single quasi-monochromatic images centered at a given wavelength while also providing a limited series of narrowband images, stepped in wavelength.

The flexibility of the TTF was well-suited for narrow-band astronomical imaging in emission lines such as [O II] 3727, [O III] 5007, Hα, [N II] 6583, [S II] 6717/6731 and [S III] 9069. Furthermore, it had the capability of obtaining images of spectral lines at arbitrary redshifts. In that sense, it also allowed for a scanning mode, but with a broader band as compared with that typically used for kinematic studies. There are several technical problems driving the development of FP Tunable Filters for narrow-band imaging within standard fixed interference filters. These problems will be highlighted ahead together with suggestions for mitigating them.

In BTFI, for the FP tunable filter, low resolution mode, we use an alternate etalon, also manufactured by SESO, deployed in the diverging beam just above the f/16.5 input focus. As explained above, SESO etalons have the capability of spanning FP gaps of ~250μm, which would, in principle allow a spectral resolution range between $500 < R < 30,000$. However, the higher end of this range is curtailed by the divergent beam, allowing resolutions of $500 < R < 2000$, for this instrumental mode.

In summary, the new interferometer can be used as a Tunable Filter (low order of interference: low resolution) and as a classical Fabry-Perot (high order of interference: high resolution) and may shift from one mode to the other very quickly. There will, inevitably be a spectral resolution "gap" between the low and high resolution FP domains which can be partially filled by the iBTF mode under certain constraints (as described in section 3.2), and by our ability to stretch the gap range of the SESO etalons themselves. This is currently under investigation.

### 3.1.3. The use of both etalons in series

As described above, the BTFI has two etalons, one in the divergent beam near the input focus (for low spectral resolutions) and the other in the collimated beam (for higher spectral resolutions). Since both etalons can be deployed independently into the optical beam, it is possible to use these etalons in tandem with the first operating as an order sorter for the second. In this manner great flexibility is achieved in the ability to select a particular order for the higher resolution etalon, a job which is usually done by the use of a fixed interference filter. By using the low resolution etalon as an order selector, avoidance of the acquisition of multiple costly interference filters (and the necessity to mount these) is achieved. The disadvantage is that order selection is achieved with a single-pass Airy profile rather than a clean near top-hat interference filter profile. Nevertheless such a two-etalon capability is seen as a significant advantage for BTFI.

### 3.2. The iBTF operating mode of BTFI

As will be described in more detail in Section 5.2, the iBTF employs two identical VPH gratings which cancel each other's dispersion. The resultant output represents the blaze function, as defined by its Bragg condition at a specific angle of incidence. By changing the angle of incidence of the grating pair this blaze function can be scanned, thus achieving wavelength tunability over a wide range of wavelengths and spectral resolutions, as defined by the grating and the range of accessible angles.

This technique gives the ability to achieve an imaging tunable filter by simply changing the angle of the grating pair; the iBTF optical configuration can employ either transmission or reflection gratings thus increasing the range of resolutions obtainable. Resolutions are then limited to those achievable with current volume phase holographic grating materials.

Gratings made from dichromated gelatin (DCG) allow for very thin grating structures with high refractive index modulations giving resolutions in the range $5 < R < 500$, while thick, low refractive index modulation gratings can be made from doped-glass which can reach resolutions towards $R \sim 4000$.

### 3.3. The Detector

When observing in wavelength scanning mode with a classical CCD, as is the case for all currently available tunable filters and Fabry-Perot systems, it can take several hours to complete a single scan and as such the observations are susceptible to changes in seeing PSF and transparency between individual frames. The result is that the profile of the scanned line will be biased, unless the scan is rapid enough so that it can then be repeated several times to achieve the desired total exposure time. The changes in seeing and transparency will be averaged when adding all the individual frames corresponding to the same scanning step of the FP (or iBTF). The problem with classical CCDs is that their readout noise will be added on each individual frame and the resulting profile will be relatively noisy. Because of the read-out noise, it is impossible to scan rapidly through the channels (one has to wait for enough counts to be collected in the frame before reading it), with the result that only observations taken in highly photometric conditions are fully reliable since the changes in seeing and transparency cannot be averaged out.

In order to be able to scan rapidly through the channels, one has to work in photon counting (or electron amplification) mode with essentially zero read-out noise. By far the best solution available today is the L3CCD (www.e2v.com). The advantage of L3 technology (an Electron Multiplying Charge Coupled Device or EMCCD) is that it operates at essentially zero read-noise as compared to a classical CCD with typically ~3 electrons of rms read-out noise. EMCCDs can be operated in an amplification mode where gain-noise imposes an effective penalty of ~2 in quantum efficiency or in photon counting mode where gain-noise can be eliminated at the cost of a serious reduction in dynamic range. For classical, long exposure, imaging or spectroscopy these disadvantages generally out-weigh the EMCCD's advantages in background-noise limited observations. However, when short exposures are demanded or when detector noise is a limiting factor, then EMCCDs can come into their own.

It will be noted that the domain of short exposure and low background noise is precisely that of the tunable filter. Not only is the background noise suppressed to a greater or lesser extent by the narrow-band imaging but the requirement to mitigate against atmospheric variability implies the use of rapid scanning whereby very short exposures are taken to build up a data-cube through continuous cycling through wavelength space. A detailed analysis showed that for BTFI, under a broad range of operating conditions, the EMCCD in amplification mode (even given the reduction of a factor of 2 in QE) gives higher signal-to-noise performance than a classical CCD when used for rapid scanning tunable filter work. Counter-intuitively, this is not the case for photon counting, despite the fact that the early use of imaging FPs used the Image Photon Counting System (IPCS). While photon counting does not suffer from gain noise inherent in the EMCCD amplification mode operation, the fact is that photon counting has such a limited dynamic range that it is only useful under the most extreme of low light level operation.

## 4. Science cases

There are a great number of galactic and extra-galactic studies which can benefit from the unique tunable filter imaging properties of the BTFI. This will be the first such instrument to work with a ground layer adaptive optics module, with a relatively large field of view (3 x 3 arcmin²). The most competitive science projects will then be those which require good spatial resolution, in particular the study of the centers of active galaxies, for investigation of the processes which drive the gas inwards and the study of stellar mass loss processes to the interstellar medium.

In July 2008 there was a survey within the Brazilian community to investigate interest in the use of such an instrument. In the extra-galactic arena the planned BTFI studies included, amongst others:

- The centers of normal and active galaxies;
- Nearby galaxies in clusters and groups;
- Mass distribution of galaxies and their building blocks;
- 2D kinematics of fine structure for galaxy modeling;
- Noncircular motions in the disks of galaxies;
- Barred galaxies;
- Kinematics of galaxies at intermediate redshift;
- Galaxy interactions and merging;
- Blue compact dwarfs, HII galaxies and tidal dwarf galaxies.

While for Galactic work and study of the interstellar medium the following topics of interest included:

- Galactic HII regions;
- Studies of Herbig-Haro objects and associated jets;
- Kinematics of Proplyds;
- Mass loss in stellar systems;
- Structure, metallicities and kinematics of planetary nebulae.

BTFI will be highly complementary to the SOAR Integral Field Spectrograph, which will also work with SAM.

## 5. BTFI Instrument Description

### 5.1. Instrument Concept

In its simplest mode the BTFI instrument is a focal reducer with a single f/16.5 collimator and dual cameras allowing the simultaneous acquisition of the filtered ($F_\lambda$) and complementary (T-$F_\lambda$) images across the observed field-of-view (T represents the spectrum of pre-filtered light incident on the tunable filter having a tunable band-pass $F_\lambda$). The simultaneous acquisition of filtered and complementary images permits a robust correction for transparency and PSF variations which otherwise plague the reconstruction of photometrically accurate 3D data-cubes.

As far as we are aware, all other FP-based imaging interferometers have done without such a facility, however accuracy of the photometric reconstruction of such data-cubes has been a severe limitation on the scientific utility of the resulting data. While high resolution kinematic data can, with care, be routinely obtained, an accurate, low resolution, tunable filter data cube requires not only superb photometric conditions over the time-frame of the spectral scan but also a stability of the image PSF to preserve spatial resolution through the data-cube. Immunity to such atmospheric instabilities can be mitigated to some degree with photon counting detectors (eg: the original TAURUS system using the Image Photon Counting System, the FaNTOMM fast scanning system or proposed systems using E2V's L3 technology). However, at low resolution, where background noise dominates, standard CCDs may still be required for ultimate sensitivity. Furthermore, the time spectrum of PSF variability as delivered by SAM's GLAO system, while it may have been modeled under the range of atmospheric conditions prevalent at SOAR, will not be confirmed until the SAM system has been fully commissioned. Hence for a system based on long time-scale sequential wavelength scanning, caution argues for inclusion of a complementary channel.

As defined above, the second, complementary, channel (T-$F_\lambda$) approximates to a continuum image of the observed field and hence offers a very deep, high signal-to-noise, image which can be used to monitor the atmosphere. However, this is not the only use of the second channel; the broad-band (T-$F_\lambda$) light can be further filtered with a FP to allow for simultaneous wavelength scanning at a secondary spectral resolution. Provided the second channel is at significantly lower resolution than the first, it can be used both as an atmospheric monitor channel and as a second science channel offering simultaneous wavelength scans at two resolutions and/or wavelengths. The two cameras of the BTFI thus represent a highly versatile instrument concept. The primary channel can be used for high resolution (FP) scans or low to intermediate resolution (iBTF) scans. In both cases the secondary channel can be used for atmospheric monitoring. Alternatively the accuracy of atmospheric monitoring can be traded with scientific utility by using the second channel for the simultaneous acquisition of data-cubes at different resolutions and/or wavelengths. The actual usage of the BTFI will be highly dependent on the science objectives of the user. The two channel concept gives the BTFI a photometric robustness for data cube acquisition while allowing a scientific versatility that is unique amongst FPs and tunable filter imagers.

### 5.2. The iBTF concept

One of the most interesting features of volume phase holographic gratings is the possibility of adjusting the efficiency curve, or "blaze" function, by varying the angle of incidence (Barden et. al. 2000a). Indeed, while their dispersive properties are identical to that of classical gratings, their diffracted energy distribution is governed by Bragg's law, as for X-rays in a crystalline structure *(*i.e. radiation that departs significantly from the Bragg condition passes through the grating undiffracted). This tunability can be advantageously used in spectrographs, but it also allows a new type of imaging tunable filter. Using a second grating, it is possible to recombine, or "undisperse", the light coming from the first grating. An image can be reconstructed as long as the gratings are parallel and have the same line fringe frequency (Blais-Ouellette et al. 2004). The iBTF concept is illustrated in Figure 2. Only light whose wavelength satisfies the Bragg condition is diffracted. It is then possible to adjust the grating angle, effectively tuning the filter's central wavelength.

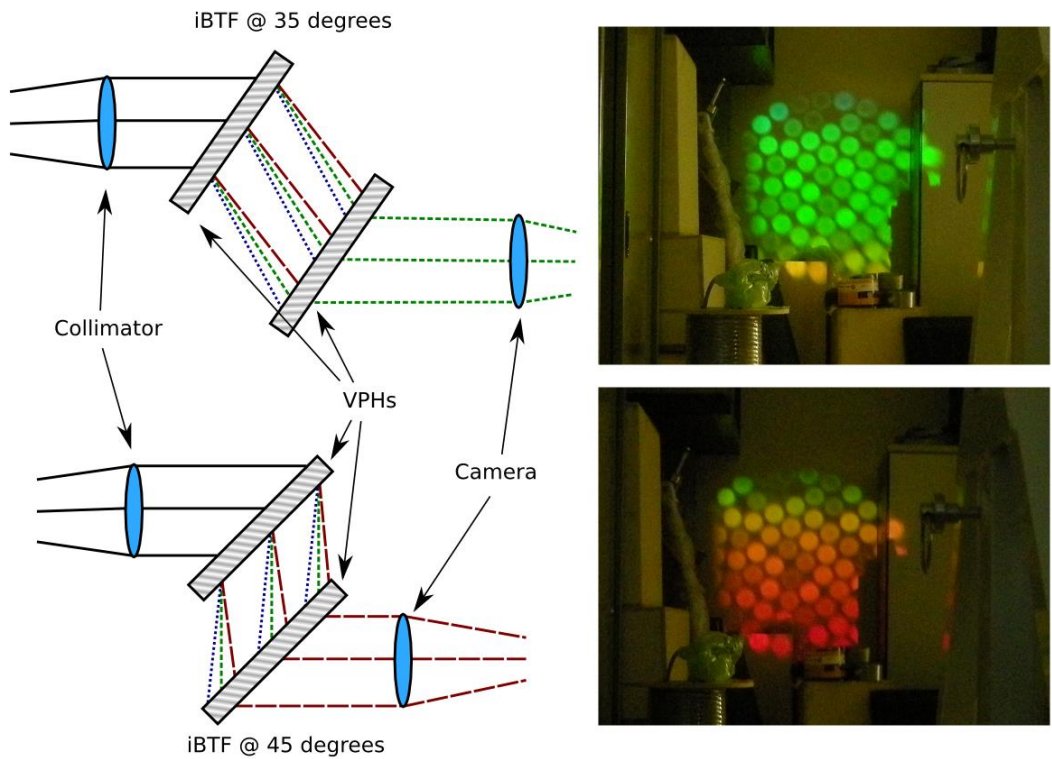

**Figure 2: This figure is a simplified representation of the dual VPH tunable filter concept. A first grating disperses collimated light that satisfies the Bragg condition. A second grating recombines the beam which is then re-imaged onto a detector. Top right: grating angle is 30º, passband is centered on 500 nm for a 2300 lines/mm VPH. Bottom right: grating angle is 45º, passband is centered on 618 nm for the same VPHs. Top and bottom left: pictures taken in the lab with the two corresponding configurations, using a common tungsten light and an optical fiber bundle as source, with a 2300 lines/mm transmission grating.**

The advantages of an iBTF tunable filter concept over a standard Fabry-Perot (or Imaging Fourier Transform Spectrograph) based instrument are as follows:
- The iBTF is compact, robust and built from custom-specified VPH gratings;
- The VPH grating is less expensive than a FP;
- Wavelength tuning is achieved through a simple rotation mechanism rather than complex and highly delicate capacitance micrometry;
- There are fewer internal alignment issues as contrasted with the highly critical and unstable plate alignment of FPs;
- Ultra-low (R > 25) together with intermediate (R ~2,000) spectral resolving powers can be routinely achieved;
- The surface of constant wavelength approximates to a $1^{st}$ order slope in the direction parallel to the dispersion axis as opposed to the complex nested paraboloids of the FP.

The only disadvantage is that it cannot achieve very high (R >4,000) spectral resolving powers if limited to standard materials and a convenient, compact format.

### 5.3. The BTFI instrument

An optical layout of the BTFI instrument is shown in Figure 3. The incident light from the f/16.5 telescope enters the instrument and is focused at the input image plane ($I_m$). The diverging beam propagates through the Field Lenses (FL) which is optionally followed by the first Fabry-Perot ($FP_{Im}$) according to the operational mode of the instrument. Afterwards, the light beam is reflected by two fold mirrors ($FM_1$ and $FM_2$) that are needed to accommodate space constraints.

The light then passes through the collimator group (CG) and, in the collimated space, it hits the first iBTF support ($GS_1$) that can hold a grating, a mirror or be empty. In the case where a grating lies in $GS_1$, the $0^{th}$ diffraction order goes straight to where the second Fabry-Perot ($FP_{Pp}$) may be. Then, it goes to the $C_1$ camera and reaches the detector $D_1$.

The $1^{st}$ diffraction order that leaves the grating at $GS_1$ goes to the second grating support $GS_2$ where the dispersion is canceled by the second twin grating. The resultant "undispersed" light is finally imaged by the $C_2$ camera at the $D_2$ detector.

The chosen 50mm pupil is compatible with readily available VPH gratings and the folds are necessary to allow the instrument to fit within the space envelope of SAM's visitor instrument port. An EMCCD detector having a format of 1600 x 1600 pixels is matched to the required pixel-scale (0.12"/pixel) and field of view (3 by 3 arcmin).

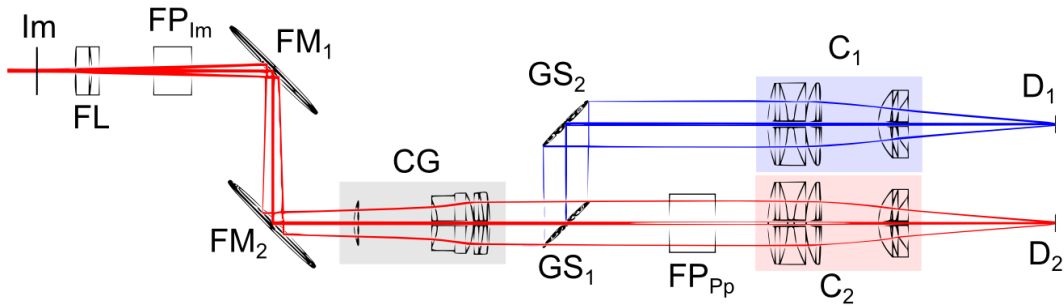
**Figure 3: Simplified representation of the BTFI optical path.**

## 5.4. BTFI performance summary

### 5.4.1. Spectral Resolving Powers

- **5 < R < 100:** The lowest resolutions can be achieved with transmission VPHGs formed from Dichromated Gelatin (DCG) (Barden et. al. 2000a);
- 100 < R < 200: In reflection the DCG gratings can deliver somewhat higher resolutions;
- 200 < R < 3,000: Using doped glass (D-G) rather than DCG intermediate resolutions can be achieved for a transmission configuration;
- 1,000 < R < 4,000: while in reflection, D-G gratings can deliver the highest iBTF resolutions attainable with current VPHG materials;
- 500 < R < 30,000: FPs (including FP-based tunable filters) are, in principle, unlimited at the high resolution end, however the R < 500 régime is very difficult to achieve in practice.

### 5.4.2. Efficiencies

VPH gratings are intrinsically very efficient gratings (~ 90% at peak). Used in double-pass they are still significantly more efficient than normal surface-ruled gratings (~60%) and FPs (~70%). In order to fully understand the relationship between the efficiency of a grating, the incident angle and the wavelength, one has to use a sophisticated theory, as the one described in (Kogelnik 1969). As an example, Figure 4 shows a simulation using this theory where one can see the efficiency curve for a specific grating (R ~ 50) in three different angles. One can also see the locus of peak efficiency as a function of wavelength which is known as the super-blaze (Barden et. al. 2000b).

Some doped glass configurations have further losses of efficiency that occur in the material itself. Nevertheless, the BTFI concept allows for a broad range of spectral resolutions at efficiencies competitive with current techniques but at a fraction of the cost and complexity.

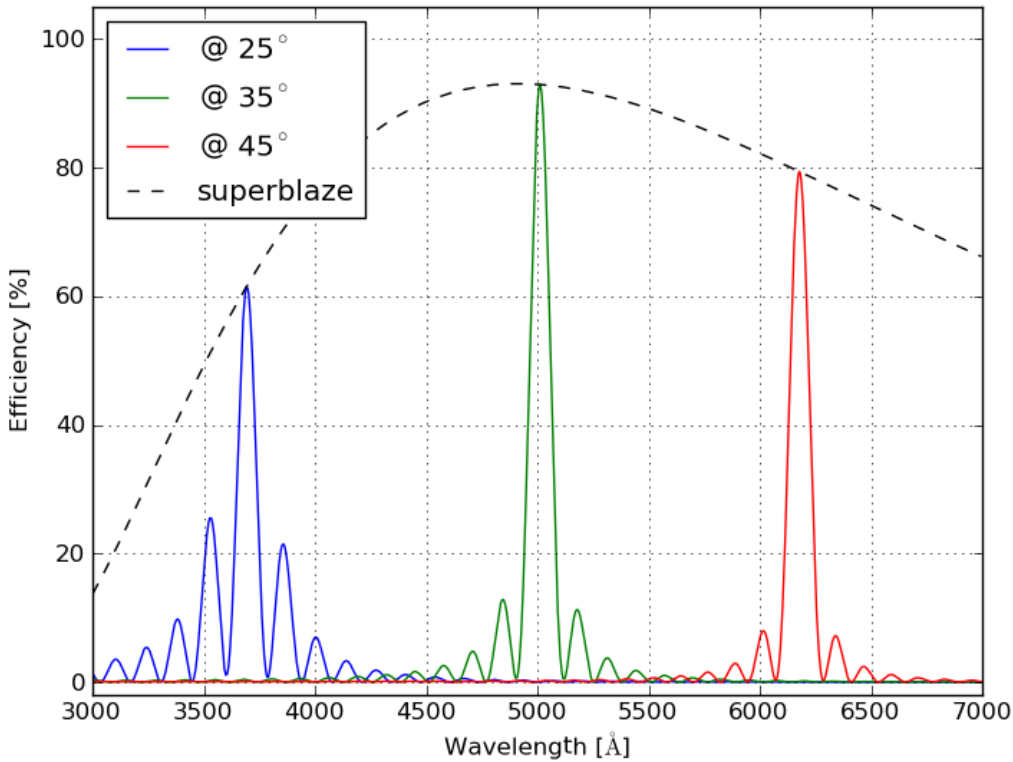

**Figure 4: Simulation using Kogelnik (1969) for a transmission grating with 2300 lines/mm, 42 um thick and refractive index modulation 0.0065, for three incident angles (solid lines) and the super-blaze function (dashed line).**

The full system throughput is a composition of SOAR, collimator and camera, fold mirrors and the EMCCD throughputs. With exception of the EMCCD, the other components have relatively flat responses with values of 0.62 for SOAR, 0.9 for the collimator and camera and 0.82 for the fold mirrors. The EMCDD throughput curve is bell-shaped with a maximum of 0.93 at 550 nm decaying to 0.25 at 350nm and 0.38 at 900nm. The full system throughput curve has the same shape with maximum at 550nm of 0.38 for SOAR+BTFI and 0.3 if SAM is included.

An acceptance test was done with the collimator and camera. The collimator was reversed and fed with a 50mm diameter collimated beam at a wavelength of 633nm. Field angles up to 2.9 degrees were tested. The back focal distance from the field lens was adjusted for best focus. The system is diffraction-limited at the best focus. Image spots were indistinguishable from an Airy disk of diameter ~0.025mm.

A similar test was done with the camera. In this case, the system was near diffraction-limited at best focus. 80% encircled energies at all field positions were a little larger than the Airy disk whose diameter was ~0.010mm.

## 5.5. Calibration and Data Reduction

There are special issues involved in the use of a FP which may have a variable gap range of < 250μm such as is the case for the SESO etalons. In particular, the SESO FP does not have a *nominal* gap from which many etalon parameters can be derived; the gap itself has to be calibrated before any particular gap setting (and hence resolution) can be established. Furthermore, because of the compactness of the BTFI instrument layout and the requirement to operate the instrument remotely, there is no easy way in which the etalons can be inspected by eye when deployed in the instrument. This presents interesting challenges to the problem of aligning, calibrating and operating the instrument. Furthermore, different procedures are required when the FP is operated in diverging and collimated beams; both configurations will be discussed.

In order to establish parallelism, four independent measures of the gap need to be established in the 4 cardinal directions across the surface of the etalon plates.

1. If the etalon is in the diverging beam then the etalon plates themselves are approximately confocal with the detector. In this case 4 spots can be illuminated with fibers, distributed in 4 cardinal points at the input focal plane.

2. If the etalon is in the collimated beam then the pupil has to be segmented so that illumination of 4 cardinal points in the pupil can be isolated. For BTFI, this is achieved using 4 small prisms near the pupil plane.

In either case, a measurement of the gap in the four cardinal positions can be achieved by illumination with two relatively nearby wavelengths ($\lambda_1$ and $\lambda_2$ with interference orders $m_1$ and $m_2$). In this case we will see that:

$$m_1 - m_2 = \lambda_1/\lambda_2 . (z_2-z_1)/\Delta z_2 - m_1 . (\lambda_1 - \lambda_2)/\lambda_2$$

where $z_1$ and $z_2$ are the emission peaks of the two wavelengths $\lambda_1$ and $\lambda_2$ as measured in etalon control units (capacitance measures proportional to gap) and $\Delta z$ is the free-spectral range, as measured in the same units. This equation can be used to determine the gap in each of the four quadrants from which both approximate parallelism and gap calibration can be achieved. Of course, fine parallelism requires the same techniques but only with one wavelength.

**Phase Calibration:** The scanning FP provides spectral line profiles for each pixel in the field. In most cases, these spectra are used simply to obtain kinematic information from the Doppler shift of a given line. The calibration process is here quite simple, since one just needs to scan a reference line, the position of which is then compared with that of the observed line at each pixel (the observed line being selected through any standard interference filter). The detailed process has been described in Amram et al. (1995). With the FP in the divergent beam, the phase map is approximately flat and hence the raw data-cube approximates to that of a series of monochromatic images. With the FP classically mounted in the collimated beam then the phase map is now parabolic and phase-correction amounts to rectifying the non-monotonic raw data-cube into a monochromatic form.

**Data reduction:** One example of a data reduction package used for FP data is that developed by Jacques Boulesteix, called ADHOC (http://www.oamp.fr/adhoc/adhocw.htm). This package is used by several groups observing with scanning FPs (e.g. IAG/USP, Brazil; UNAM, Mexico; Observatoire de Paris Meudon; Université de Montréal and Observatoire du Mont Mégantic, Québec; Byurakan Observatory, Armenia; SAO Zelenchuk, Russia). Illustrations of the data reduction process can be found at the following link: http://www.oamp.fr/PdG/GHASP/ghasp_en.htm.

### 5.6. EMCCD Cameras

As stated above, EMCCD detectors offer images free of read-out noise. Read-out noise is added by the output amplifier at the very last stage of the detector, where the charge in electrons is converted to a measurable voltage. Thanks to an electron multiplication process that occurs before reaching the output amplifier in EMCCDs, each electron (created from incoming photons) generates thousands of electrons. This is a stochastic process producing the ~2 reduction in QE mentioned above. The read-out noise added by the output amplifier is still present, but its effect is greatly diminished to negligible values. This is generally stated as sub-electron readout noise (Daigle et al. 2009).

However, any noise added to the photoelectrons before the multiplication process will suffer its multiplicative effect. An example of such a noise source are cosmic rays.

The main noise source in EMCCD that depend on the detector readout electronics is known as Clock Induced Charge (CIC) or 'spurious charge' (Tulloch, 2005) and is generated by the transitions of the voltage phases used to transfer the electrons across the device when storing or reading out an image. There are various methods to reduce CIC noise, usually employing wave-shape in the phases (Janesick 2001, so the effect of the transitions is minimized) and running the device in non-inverted mode of operation, or non-MPP, Multi-Phased Pinned (for more details see e2v Technical Note 4, 2004). Unfortunately, dark current is greatly increased in non-MPP mode.

We decided to design and build our EMCCD cameras, as opposed to using commercially available cameras, in order to benefit from a carefully defined set of requirements, such as deep-cooling (lower than -100 C) and arbitrary clocking. We are using readout electronics from the University of Montreal which were specially designed for EMCCDs and are now being commercialized by Nüvü Caméras (www.nuvucameras.com, Daigle et al. 2009). The BTFI cameras were built in collaboration with Universidad Catolica in Chile.

## 5.7. Current Status

The iBTF concept has been successfully prototyped through an NSF grant (Award #0352991) confirming the basic functionality and applicability of the double-pass VPHG concept both in transmission and reflection modes. This prototype has now been developed into a commercial product (by the company Photon etc, Montreal) as a laboratory tunable narrow band source and spectrophotometer for instrument and filter calibration. However, there are no other references regarding the use of twin VPHG's as tunable filters, given that BTFI will be the first instrument of this kind in Astronomy. For BTFI we have first developed a schematic optical layout (Figure 3) which satisfied the space constraints of SAM's visitor instrument port and we have developed an opto-mechanical design as shown in Figure 5. In September, 2007, the BTFI project successfully passed through its Concept Design Review and in June, 2008, the Preliminary Design review, the panel members of which were selected from the international astronomical instrument community.

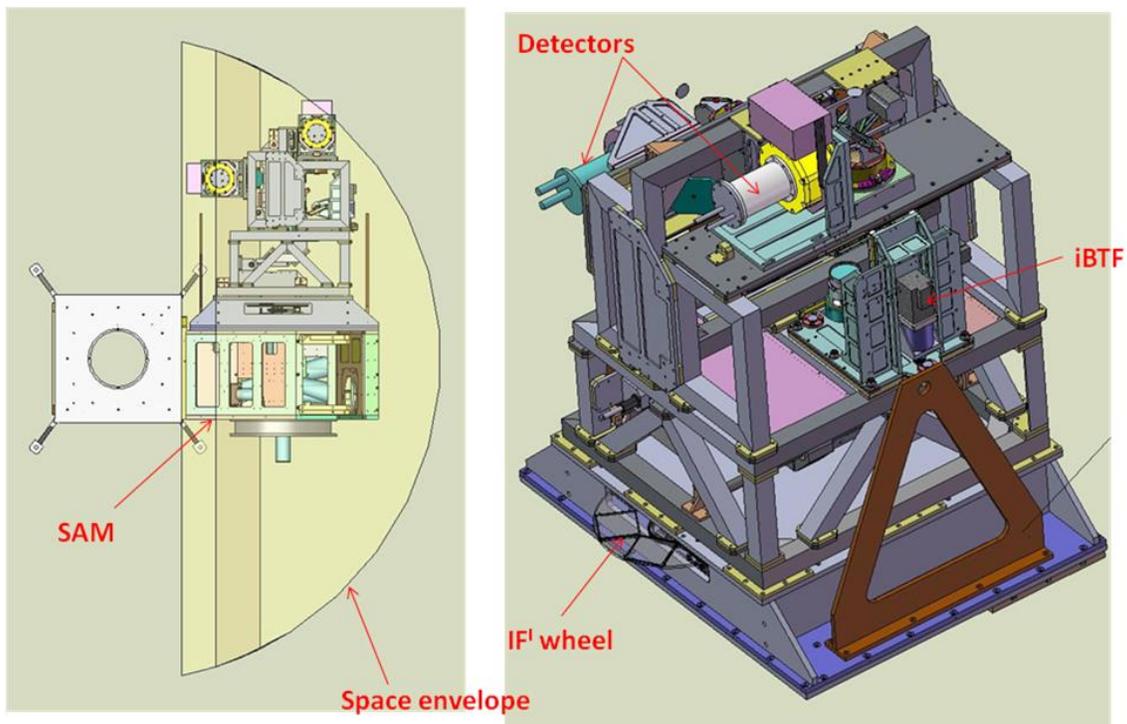

**Figure 5: An overview of the instrument mechanical design. Finite element analysis has validated the structural design.**

The total cost of the instrument is ~ $1.5 million (USD) and was mostly funded through grants from FAPESP, the Research Funding Agency of the State of São Paulo, supplemented by additional funds from LNA and the Conselho Nacional de Pesquisa

(CNPq). This includes hardware and contract labor, excluding substantial in-house support using IAG and INPE labor.

The project is now in its commissioning phase as a visitor instrument at SOAR. Final stages of electro-mechanical integration took place at USP and INPE in 2010A. It was mounted on the direct port of the SOAR telescope and had its first light in 2010B.

An example data obtained with BTFI is shown in Figure 6. Both images are taken from a data-cube resulting from the observation of the planetary nebulae NGC 2440 using the iBTF with a reflection grating with 2370 lines/mm and scanning from 38.75º to 41.00º in steps of 0.05º.

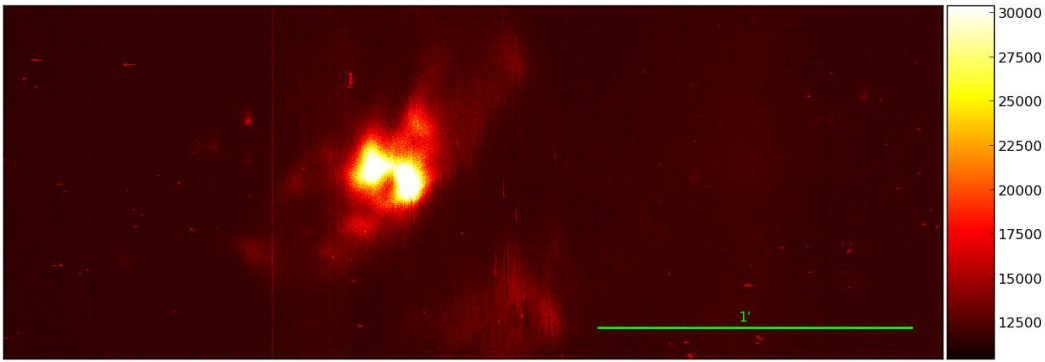

**Figure 6: Data-cube obtained from NGC 2440 collapsed in λ.**

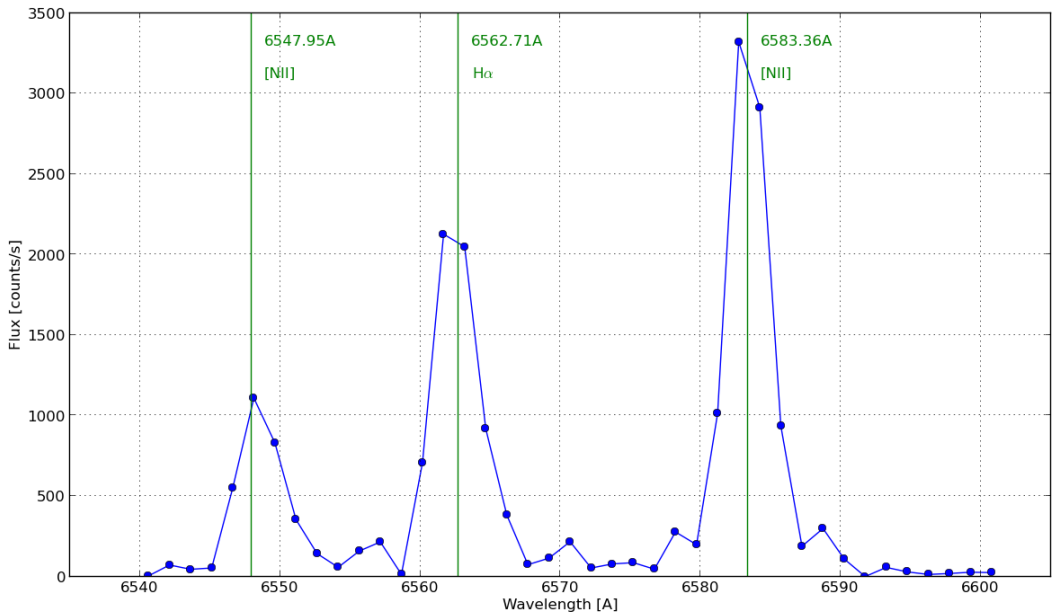

**Figure 7: Spectrum extracted from the central pixel of the upper-left lob of the NGC 2440.**

6. **Conclusion**

The Brazilian Tunable Filter Imager represents an instrument strategy that optimizes the science potential for optical spectroscopy with the SOAR telescope, with its emphasis on high image quality and its use of Ground Layer Adaptive Optics for image enhancement in the optical over a field of view of 3 x 3 arcmin. It is expected that it will become a regular users' instrument after SAM is fully commissioned.

The authors thank FAPESP (grant 2006/56213-9) and the Instituto Nacional de Ciência e Tecnologia de Astrofísica (grant 2008/08/57807-5), CNPq and LNA/MCTI for funding of the BTFI project. Fellowships granted to four Masters students, one PhD student, and four technicians (technical capacitation fellowships) who worked in the project are acknowledged to FAPESP, CAPES and CNPq. The BTFI team thanks the CTIO technical team for their valuable support and the SOAR director Steve Heathcote, without whom this instrument would have never been completed.

Hernandez, O., Gach, J., Carignan, C. & Boulesteix, J. 2003, SPIE, 4841, 1472

Ingraham, P., Doyon, R., Beaulieu, M., Rowlands, N. & Scott, A. 2010, SPIE 7731, 121

Janesick, J. R. 2001, Scientific Charge-Coupled Devices (SPIE Press Monograph; PM 83. ISBN 0819436984; Bellingham, WA: SPIE Optical Engineering Press), 651.

Kogelnik, H., 1969 Coupled wave theory for thick hologram gratings, The Bell System Technical Journal, Vol. 48, no. 9, p. 2909
Le Fèvre et al, 1994A&A...282..325L

Lepine, J. R. D., de Oliveira, A. C., Figueredo, M. V., Castilho, B. V., Gneiding, C., Barbuy, B., Jones, D. J.,Kanaan, A., Mendes de Oliveira, C., Strauss, C., Rodrigues, F., Andrade, C. R., de Oliveira, L. S., de Oliveira, J. B., 2003, SPIE, 4841, 1086.

Rangwala, Naseem; Williams, T. B.; Pietraszewski, Chris; Joseph, Charles L. 2008, AJ, 135, 1825

Roesler, F. L. and Mack, J. E. 1967: J. Physique 28, Supp. C2, 313 - 320.

Rosado, M.; Langarica, R.; Bernal, A.; Cobos, F.; Garfias, F.; Gutierrez, L.; Tejada, C.; Tinoco, S.; Le Coarer, E. 1995, RMxAC, 3, 263

Smith, W. H. 1981, in Mod. Observational Tech. for Comets, p. 156.

Taylor, K.; Atherton, P. D. 1980: Monthly Notices of the Royal Astronomical Society, vol. 191, p. 675-684.

Tokovinin, A., Tighe, R., Schurter, P., Cantarutti, R., van der Bliek, N., Martinez, M., Mondaca, E., Montane, A. 2008, SPIE, 7015, 70154.

Tulloch, S., Scientific Detectors for Astronomy, 2005, 303.

Tully, R. Brent 1974: Astrophysical Journal Supplement, vol. 27, p.415

Veilleux, S.; Weiner, B. J.; Rupke, D. S. N.; McDonald, M.; Birk, C.; Bland-Hawthorn, J.; Dressler, A.; Hare, T.; Osip, D.; Pietraszewski, C.; Vogel, S. N. 2010, AJ, 139, 145-157.

de Vaucouleurs, G.; Pence, W. D. 1980: Astrophysical Journal, vol. 242, p. 18-29.

Williams, T. B.; Caldwell, N.; Schommer, R. A. 1984: Astrophysical Journal, vol. 281, p. 579-584.